\documentclass[3p, twocolumn]{elsarticle} 
\usepackage{graphicx}
\usepackage{listings}
\usepackage{natmove}

\journal{Computational Materials Science}

\begin{document}

\begin{frontmatter}

  \title{MPInterfaces: A Materials Project based Python Tool for
    High-Throughput Computational Screening of Interfacial Systems}

  \author[cornell,uf]{Kiran Mathew\corref{cor1}}
  \ead{km468@cornell.edu}
  \author[nist]{Arunima K. Singh}
  \author[uf]{Joshua J. Gabriel}
  \author[uf]{Kamal Choudhary}
  \author[penn]{Susan B. Sinnott}    
  \author[nist]{Albert V. Davydov}
  \author[nist]{Francesca Tavazza}
  \author[uf]{Richard G. Hennig\corref{cor1}}
  \ead{rhennig@ufl.edu}

  \cortext[cor1]{Corresponding author}
  \address[cornell]{Department of Materials Science and Engineering,
    Cornell University,
    Ithaca, NY 14850, U.S.A.}
  \address[nist]{Materials Science and Engineering Division,
    National Institute of Standards and Technology,
    Gaithersburg, MD 20899, U.S.A. }  
  \address[uf]{Materials Science and Engineering,
    University of Florida,
    Gainesville, FL 32611, U.S.A. }
  \address[penn]{Department of Materials Science and Engineering,
    The Pennsylvania State University,
    University Park, PA 16801, U.S.A}

\begin{abstract}
  A Materials Project based open-source Python tool,
  \textit{MPInterfaces}, has been developed to automate the
  high-throughput computational screening and study of interfacial
  systems. The framework encompasses creation and manipulation of
  interface structures for solid/solid hetero-structures,
  solid/implicit solvents systems, nanoparticle/ligands systems; and
  the creation of simple system-agnostic workflows for in depth
  computational analysis using density-functional theory or empirical
  energy models. The package leverages existing open-source
  high-throughput tools and extends their capabilities towards the
  understanding of interfacial systems. We describe the various
  algorithms and methods implemented in the package. Using several
  test cases, we demonstrate how the package enables high-throughput
  computational screening of advanced materials, directly contributing
  to the Materials Genome Initiative (MGI), which aims to accelerate
  the discovery, development, and deployment of new materials.
\end{abstract}


\end{frontmatter}


\section{Introduction}
\label{sec:mpint_intro}

Interfaces play a vital role in practically all materials and
devices~\cite{Joo2003, Park2004, Bera2010, Ha2014}. For instance, the
efficiency and stability of electrochemical devices are mostly decided
by the composition and the properties of the solid electrolyte
interface layers~\cite{Lu2014}. In another example, the self assembly
of nanoparticles used in high-efficiency photovoltaic devices is
directed by the interfaces formed between nanoparticles, ligands and
the solvent they are dispersed in~\cite{Baumgardner2013,
  bealing12}. As materials and devices are getting smaller, the
interface properties begin to dominate their essential
characteristics. Progress in a wide variety of applications ranging
from catalysis to microelectronics is guided by a refinement of our
understanding and control of the interface properties.

Experimental studies combined with computational investigations
provide a broad spectrum of information needed for an accurate and
thorough understanding of interfaces, which is seldom accessible by
utilizing one of these two approaches alone~\cite{Ceder1998, Ha2014,
  Choi2011,Hwang13, Ha2013, Anasori2015}. Though combinatorial
techniques are increasingly used by experimentalists for identifying
new compositions as well as for the rapid optimization and mapping of
processing parameters that influence the properties of materials, a
complete experimental characterization of a large number of possible
candidate systems poses a daunting challenge due to time-consuming and
expensive experiments, besides limitations involved while exploring
extreme and hazardous environmental conditions~\cite{vanDover1998,
  Koinuma2004, Takeuchi200518}. In order to complement and guide
experimental investigations, and to accelerate the discovery of novel
phenomenon at interfaces it is imperative to adopt a rational approach
towards the screening and the characterization of interfacial
systems. The advancement of modern computing has pushed the boundaries
of materials simulations making them faster, more cost-effective,
efficient and accurate. In recent years, high-throughput computational
studies have predicted new materials for applications in
photocatalysis, energy storage, piezoelectrics and electrocatalysis
~\cite{Singh2015, Cheng2015, Armiento2011, Greeley2009,
  andersson2006toward, greeley2006computational,
  Greeley2009nchem}.

The study of materials interfaces presents additional challenges
compared to bulk materials' properties due to the increased number of
configurations and conformations possible in these two (or more) phase
systems and requires larger computational resources due to the reduced
symmetry of the system. To enable the high-throughput computational
screening of the structure, stability, and properties of materials
interfaces -- such as between nanocrystals, ligands, and solvents and
between 2D materials and substrates -- we have developed the
open-source Python package, \textit{MPInterfaces}. First, the package
automates the generation of various interfacial structures and
prepares input files to first principles density-functional theory
(DFT) simulations softwares like Vienna Ab-initio Software Package
(VASP)~\cite{vasp} and molecular dynamics (MD) simulations softwares
like Large-scale Atomic/Molecular Massively Parallel Simulator
(LAMMPS)~\cite{Plimpton1995}. It then enables the creation of
high-throughput computational workflows that can be deployed on remote
computing resources. Finally, the package provides analysis tools such
as for the prediction of the shape of nanocrystals using surface
energies and the Wulff construction. The coupling with the energy
calculation softwares and the workflow creation builds on the
framework of the open source Python packages of the Materials Project
namely pymatgen~\cite{Ong2012b}, custodian~\cite{Ong2012b} and
fireworks~\cite{fireworks}.

The package is being continuously developed and the latest version can
be obtained from the GitHub repository at
\url{https://github.com/henniggroup/MPInterfaces}. In the following
sections we present an overview of the package using several examples,
describe its capabilities, and discuss the algorithms we employed to
overcome some of the technical and scientific challenges.

\section{MPInterfaces: Overview with examples}
\label{sec:mpint_overview}

The \textit{MPInterfaces} package is written in Python 2.7 with
support for Python 3.x. The package makes extensive use of existing
Python tools for the generation and manipulation of various structures
and the creation of the corresponding input files for DFT and MD
simulations.  For the structural analysis and input file generation
for the interfacial structures we extend the pymatgen
package~\cite{Ong2012b}. The Atomic Simulation
Environment~\cite{ASE2002} package is used to interface with the MD
software package LAMMPS. For the workflow creation and management we
extend the custodian and the fireworks packages~\cite{Ong2012b,
  fireworks}. In the following subsections we discuss the capabilities
supported by the \textit{MPInterfaces} package that are built on top
of the aforementioned packages. Illustrative examples with
corresponding code snippets are provided in each section to aid users
in employing this framework.

\subsection{Ligand capped nanoparticles}
\label{sec:mpint_ligand_nano}

Nanocrystals in the form of quantum dots are increasingly employed in
the fabrication of devices such as solar cells, transistors, and
LEDs~\cite{Dey2013, Bohn2012}.  Often the assembly of such
nanocrystals into superlattices yields new materials with tunable
optical and electronic properties~\cite{Baumgardner2013,
  bealing12}. DFT calculations and MD simulations are routinely used
to characterize interfaces properties that are hard to access with
stand alone experiments, such as for example the binding energies and
surface energy changes in varying environments of capping ligand and
solvents~\cite{Choi2011, bealing12, Fishman2013}.

It is well known that nanocrystal surface chemistry as well their
shape evolution are controlled by the thermodynamics and kinetics of
the three phase system comprising of the nanocrystals, surfactant
molecules (also known as capping ligands) and the
solvent~\cite{Choi2011}. Furthermore, the energetics of the system
determines the thermodynamics and kinetic barriers of the growth of
such nanocrystals, which drive their size, shape, and properties.

\begin{figure}[t!]
  \includegraphics[width=\columnwidth]{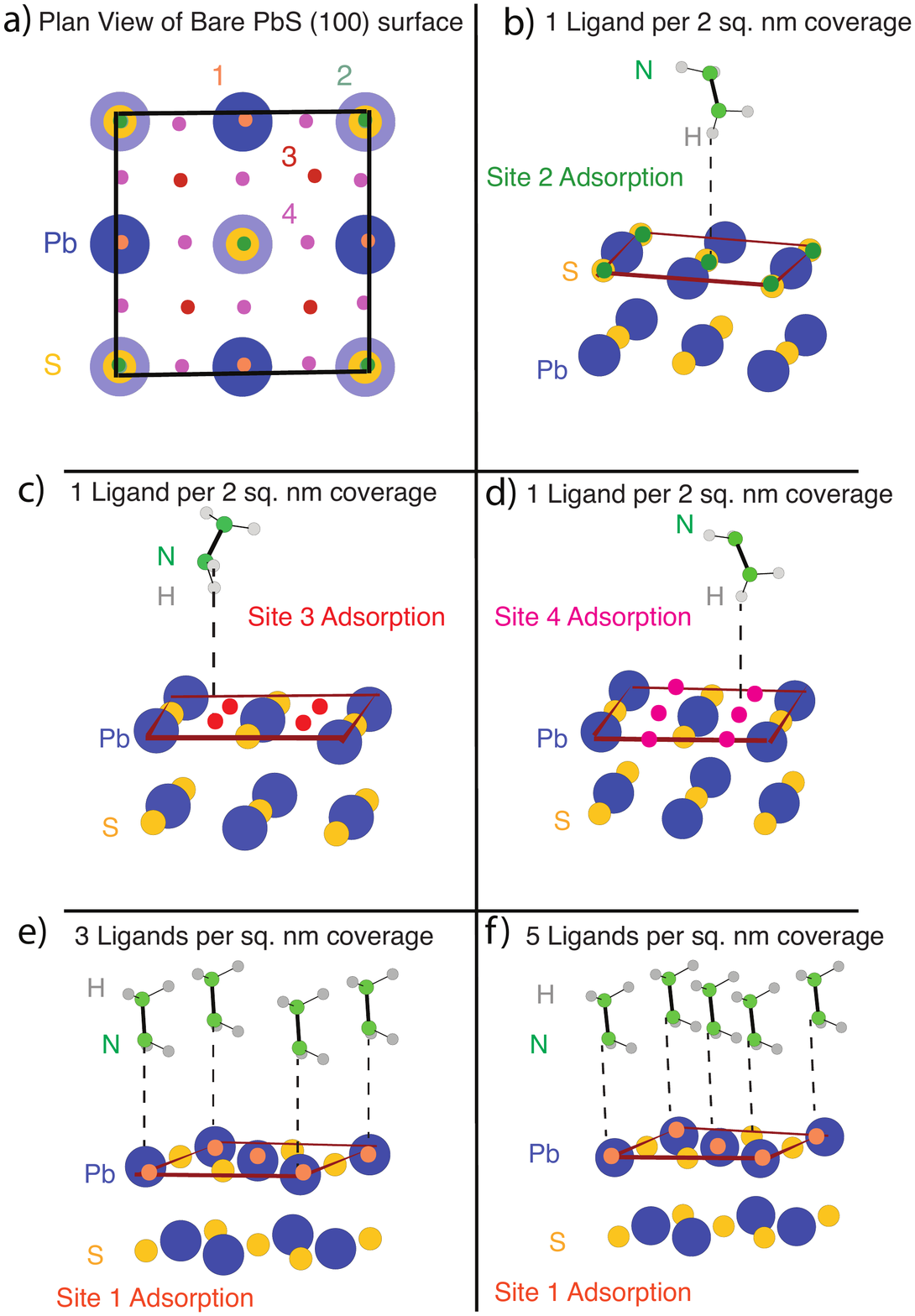}
  \caption{ \label{fig:Ligand Nanocrystal Interface} Specification of
    the structure of ligands adsorbed on PbS(100) facet. (a) Top view
    of a rocksalt PbS(100) facet with possible high-symmetry binding
    sites 1,2,3,4 marked as orange, green, purple and red circles,
    respectively. The Pb atoms are shown in blue and S in yellow.  (b)
    A hydrazine ligand is placed on site 2, above a S atom bound
    through the H atom of the ligand, (c) site 3 of PbS(100)
    referenced with adsorption distance bound through the Hydrogen (H
    - Grey) atom on the ligand, (d) site 4 of PbS(100) bound through
    the N atom on the ligand with a random rotation about cartesian
    axes, (e) higher coverage on site 1 above a Pb atom, binding
    through N atom, (f) a high coverage over site 4, binding through N
    atom. Notice the preservation of symmetry between the molecules to
    minimize the steric hindrance. Other entropic configurations can
    be included with manual manipulation of binding sites.}
\end{figure}

\textit{MPInterfaces} enables a high-throughput computational screening of
nanoparticle/ligand combinations by sampling the following degrees of
freedom in the system: (a) the crystallographic planes of the surface
facets, (b) the ligand binding sites on each facet, and (c) the ligand
surface coverage for the given facet-ligand-binding site
combinations. We model these interfaces using slab models of a
specified thickness and crystallographic facet, and with ligands
adsorbed on the surfaces. The construction of the slab model for an
interface requires the specification of the following six parameters:
(i) the bulk phase that constitutes the surface in question, (ii) the
crystallographic facet of interest in this bulk phase, (iii) the
initial binding site on the surface, (iv) the binding atom on the
ligand that is intuitively expected to be the nearest neighbor to the
chosen binding site on the surface, (v) the initial approximate
separation between the surface binding site and the atom, which in
other words is the adsorption distance, and (vi) the ligand coverage
on the surface in number of ligands per unit
area. Figure~\ref{fig:Ligand Nanocrystal Interface} illustrates as an
example the degrees of freedom that can be controlled for an interface
model of a (100) surface of PbS that is decorated with lead acetate,
Pb(CH$_3$COO)$_2$, ligands.

The \textit{interfaces.py} module in the package defines the Python
classes for the creation of ligands from the combinations of different
molecules.  The molecule structures can be generated using the Python
interface to openbabel~\cite{OpenBabel} or directly read in from the
locally available structure files of various formats supported by
pymatgen. The ligands are then placed above a slab that is generated
from a bulk structure by specifying the \textit{(hkl)} Miller indices
of the required facet. Figure~\ref{fig:Ligand Nanocrystal Interface}
illustrates how the ligand configurations are specified through the
binding sites and surface coverage. Additionally a liquid phase, such
as a solvent or electrolyte, can be added to the nanoparticle-ligand
interface in an efficient and accurate manner by using the implicit
solvent or electrolyte models that are provided by the VASPsol
module~\cite{Mathew2014, Mathew-arxiv2016}.

The following code excerpt illustrates how the slab structure in
Fig.~\ref{fig:Ligand Nanocrystal Interface}(e) is generated in the
\textit{MPInterfaces} framework.
\begin{lstlisting}
  # A code excerpt for generation of a single 
  # nanoparticle facet-ligand interface
  # All distances are in Angstroms
  
  from mpinterfaces.interface import Ligand, \
                                     Interface
  
  Bulk_struct = Structure.from_file("POSCAR.bulk")
  hkl = [1,0,0]
  min_thick = 10
  min_vac = 30
  # ligands per sq. Angstroms
  surface_coverage = 0.03
  hydrazine = Molecule.from_file(
                            "hydrazine.xyz")
  ligand= Ligand([hydrazine])
  # position the ligand
  x_shift = 0.0
  y_shift = 0.0
  z_shift = 3.0
  adsorb_on_species = 'Pb'
  adatom_on_ligand='N'  
  interface = Interface(
           bulk_struct, hkl=hkl,
           min_thick=min_thick,
           min_vac=min_vac,
           ligand=ligand,
           displacement=z_shift,
           x_shift=x_shift, y_shift = y_shift           
           adatom_on_lig=adatom_on_ligand, 
           adsorb_on_species= adsorb_on_species,
           surface_coverage=surface_coverage)                    
  interface.create_interface()
\end{lstlisting}  
This code can be easily generalized to create and simulate arbitrary
facets of bulk materials with atomic or molecular species adsorbed on
the surface.

\subsection{Wulff construction}
\label{sec:mpint_wulff}

As mentioned above, the shape of nanocrystals as well as their
self-assembly into mesoscale structures depends strongly on the
thermodynamics and kinetics of the system, comprised of the
nanoparticles, ligands and solvent. The shape of a crystal or
nanocrystal is give by the Wulff construction, which requires as input
the surface energy of the crystal facets \cite{Wulff1901,Fonseca125,
  Kim2009, Heyraud1983}.

For nanocrystal surfaces capped with ligands, the surface energy,
$\gamma_{hkl}$, is a function of the ligand coverage and given
by~\cite{bealing12}
\begin{displaymath}
  \gamma_{hkl} = \gamma^0_{hkl} - \Theta_{hkl} E_{b,hkl},
\end{displaymath}
where $\gamma^0_{hkl}$ is the surface energy of the bare slab,
$\Theta_{hkl}$ is the surface coverage, and $E_{b,hkl}$ is the binding
energy of the ligand(positive values indicate strong binding). Given
the surface energies of all facets of interest in the crystal
structure, it is possible to predict the equilibrium shape of a
nanocrystal, which is the multifaceted shape that minimizes the
nanocrystal's surface energy. This shape can be obtained using the
Wulff construction. Our implementation of the Wulff construction
extends existing implementations for cubic symmetry to arbitrary space
groups. The code takes advantage of the space group symmetry and
utilizes the pymatgen symmetry tools to determine the
symmetry-equivalent crystallographic facets.

\begin{figure}
  \centerline{\includegraphics[width=6cm]{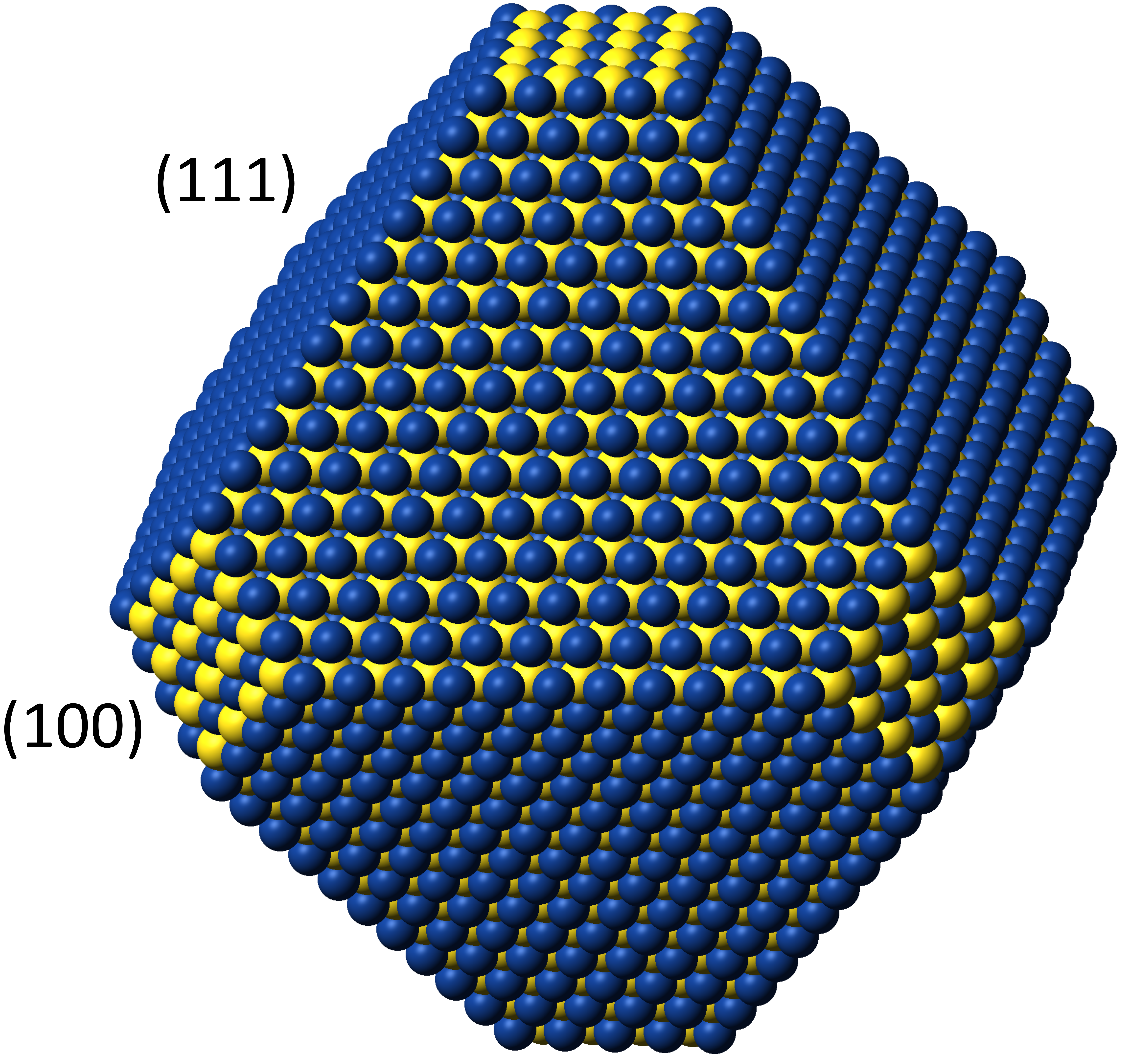}}
  \caption{ \label{fig:pbs_tmpo} The shape of a PbS nanocrystal
    generated using the Wulff construction using the
    \textit{MPInterfaces} code. The code excerpt used for the creation of
    nanoparticle is provided in the text. The PbS nanocrystal exhibits
    the shape of a truncated octahedron with (111) and (100) facets.
  }
\end{figure}

The \textit{nanoparticle.py} module in the \textit{MPInterfaces}
package implements the classes that let the user provide the list of
Miller index families, their corresponding surface energies, and the
maximum radius of the nanoparticle to create the equilibrium
nanocrystal shape using to the Wulff construction. The following code
excerpt illustrates the construction of the equilibrium shape of the
PbS nanocrystal shown in Figure~\ref{fig:pbs_tmpo}. The surface
energies of the three low-index (111), (100), and (110) facets of PbS
in vacuum are taken from Ref.~\cite{Mathew2014}. The surface energy of
the (111) facet is for the reconstructed surface~\cite{bealing12},
however, we do not show the (111) surface reconstruction in
Figure~\ref{fig:pbs_tmpo}. The resulting nanocrystal shape is a
truncated octahedron formed by (111) and (100) facets.

\begin{lstlisting}
  # A code excerpt for generation of
  # PbS nanoparticle
  # All distances are in Angstroms and
  # surface energies in meV per square
  # Angstroms
  
  from mpinterfaces.nanoparticle import \
                                 Nanoparticle

  rmax = 50
  hkl_family = [(1,1,1), (1,0,0), (1,1,0)]
  surface_energies = [18, 24, 28]
  nanoparticle = Nanoparticle(
                 bulk_structure,
                 rmax=rmax,
                 hkl_family=hkl_family,
                 surface_energies=surface_energies)
  nanoparticle.create()
\end{lstlisting}  

\subsection{Heterostructure Interfaces}
\label{sec:mpint_hetero}

The first step in the study of heterojunctions of solid-state
materials is the construction of the interface between two crystal
structures. This construction is challenging due to the fact
that there are usually numerous possible coincident site lattices
(CSLs)~\cite{Santoro1973, Hwang13}, which can be used to create the
heterostructure interface. Moreover, additional possibilities arise
due to all the possible placement of atoms at the interface. A very
simplistic approach would to be to do a brute force search of the
coincidence lattice vectors between two surfaces. However this is not
only inefficient but is also inapplicable for surfaces with different
crystal symmetry~\cite{Hwang13}.

\begin{figure}[t]
  \begin{center}
    \begin{tabular}{c}
      \includegraphics[width=\columnwidth]{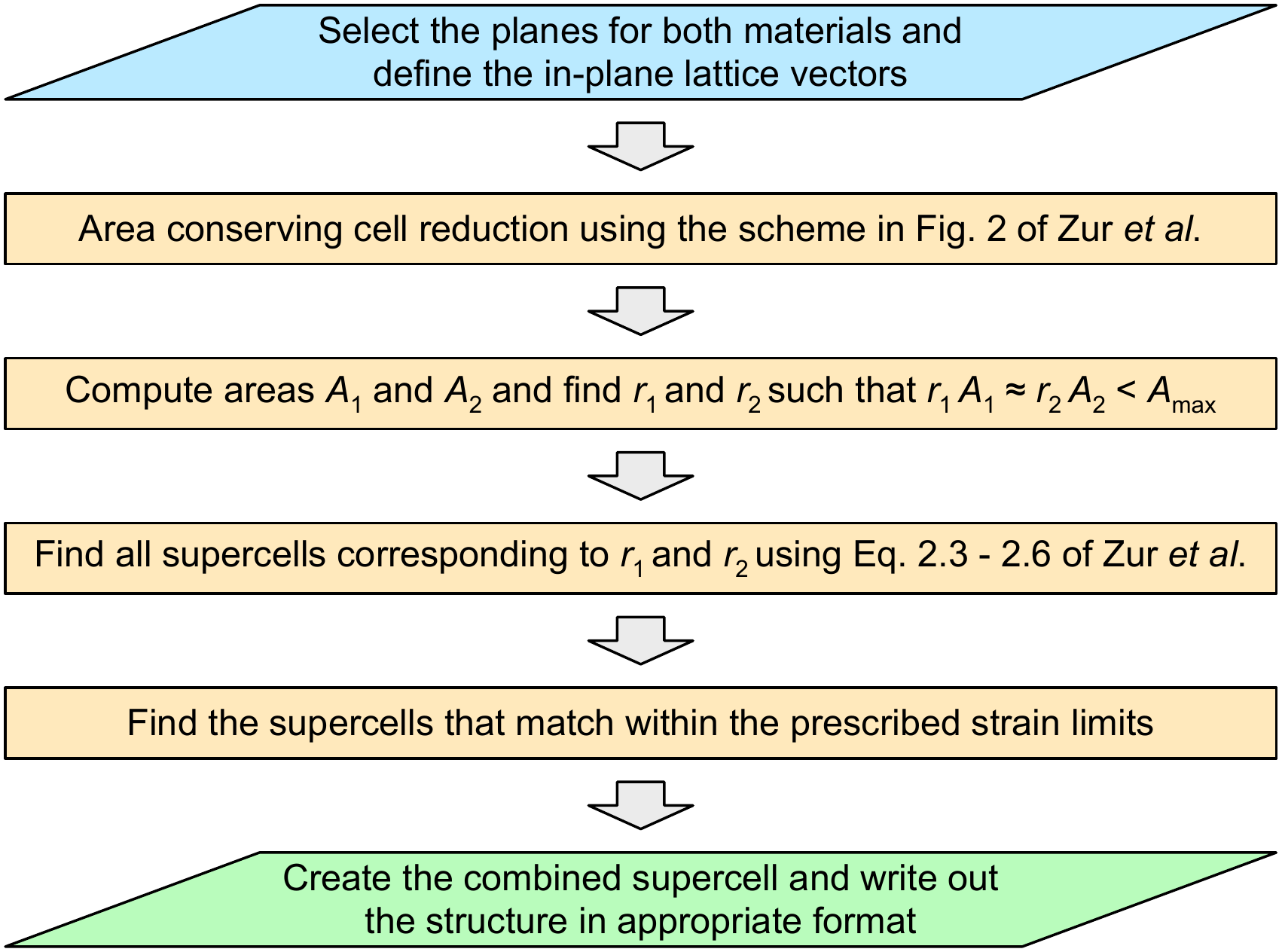}
    \end{tabular}
  \end{center}
  \caption[Zur] 
          { \label{fig:Zur} The algorithm of Zur
            $et~al.$~\cite{zur1984lattice} for the epitaxial
            lattice-matching and symmetry-matching of any two surfaces
            for all rotations; within a given lattice mismatch and
            within a given surface area of the interface. }
\end{figure}

In the \textit{MPInterfaces} package we implemented an algorithm,
which rapidly scans through various interface configurations for an
interface between two crystal surfaces and identifies those pairs that
are within a specified lattice-mismatch, symmetry-matched, and
distinct from each other.  Figure~\ref{fig:Zur} illustrates the
efficient lattice matching algorithm proposed by Zur {\it et
  al.}~\cite{zur1984lattice}, which we employ to identify the lattice
and symmetry matched interfaces for any two crystal surfaces and
arbitrary rotations. First, the number of possible CSLs in this
algorithm depends on the maximum permitted interface area and the
maximum lattice-mismatch that the interface is allowed to
undergo. Second, the surface matching of the interface atoms is
performed by identifying the distinct atoms in the near-interface
layers and creating all distinct structures formed by placing the
non-distinct atoms on top of each other.

Combined with the input file generation and workflow management
features of \textit{MPInterfaces}, this algorithm can be used to
perform high-throughput computational screening of suitable substrates
for 2D material deposition and functionalization~\cite{SinghHigh}.
Given a list of 2D materials and substrates, using the
\textit{MPinterfaces} package we find the CSLs for each substrate/2D
material pair and subsequently generate the structure files for the
combined structure for further DFT or MD analysis. A study of the
thermodynamics of the system and effects of substrate on the 2D
material properties can reveal if the growth of a particular 2D
materials is feasible on that substrate; and if the substrate alters
any of the 2D materials properties~\cite{SinghHigh, Singh2015al2o3,
  Singh2014_synth, Singh2014GaN, Zhuang2013III-V}.

\begin{table}[]%
  \caption{Identification of the interface structure for graphene on
    seven possible metal (111) substrates. The lattice parameter, $a$
    (\AA), of the (111) surfaces of Pt, Cu, Au, Pt, Ni, Al and Ag are
    shown, together with lattice-mismatch between the primitive
    lattices of graphene and the substrates, $\epsilon_{1:1}$ in (\%),
    and compared with the lattice-mismatches obtained from the
    \textit{MPInterfaces} code, $\epsilon_\mathrm{\it MPInt}$
    (\%). The supercells of substrate, $S_\mathrm{sub}$, and graphene,
    $S_\mathrm{gr}$, obtained from\textit{MPInterfaces} are provided as well.}
\label{tab:lat_mismatch}
\begin{tabular}{crrrrr}
\hline
  { } &   $a$ & $\epsilon_{1:1}$  & $\epsilon_\mathrm{\it MPInt}$ & $S_\mathrm{sub}$ & $S_\mathrm{gr}$ \\
\hline
\hline
  Pt &      3.98 &      12.55 &       0.98 & $\sqrt{3}\times\sqrt{3}$ & $2\times2$ \\
  Cu &      3.62 &       3.92 &      -3.92 & 1$\times$1 & 1$\times$1 \\
  Au &      4.17 &      16.45 &      -3.53 & $\sqrt{3}\times\sqrt{3}$ & $2\times2$ \\
  Pd &      3.95 &      11.71 &       1.94 & $\sqrt{3}\times\sqrt{3}$ & $2\times2$ \\
  Ni &      3.52 &       0.93 &      -0.93 & 1$\times$1 & 1$\times$1 \\
  Al &      4.07 &      14.37 &      -1.12 & $\sqrt{3}\times\sqrt{3}$ & $2\times2$ \\
  Ag &      4.14 &      15.82 &      -2.79 & $\sqrt{3}\times\sqrt{3}$ & $2\times2$ \\
\hline
\hline
\end{tabular}  
\end{table}

\begin{figure}[t]
  \centerline{\includegraphics[width=7cm]{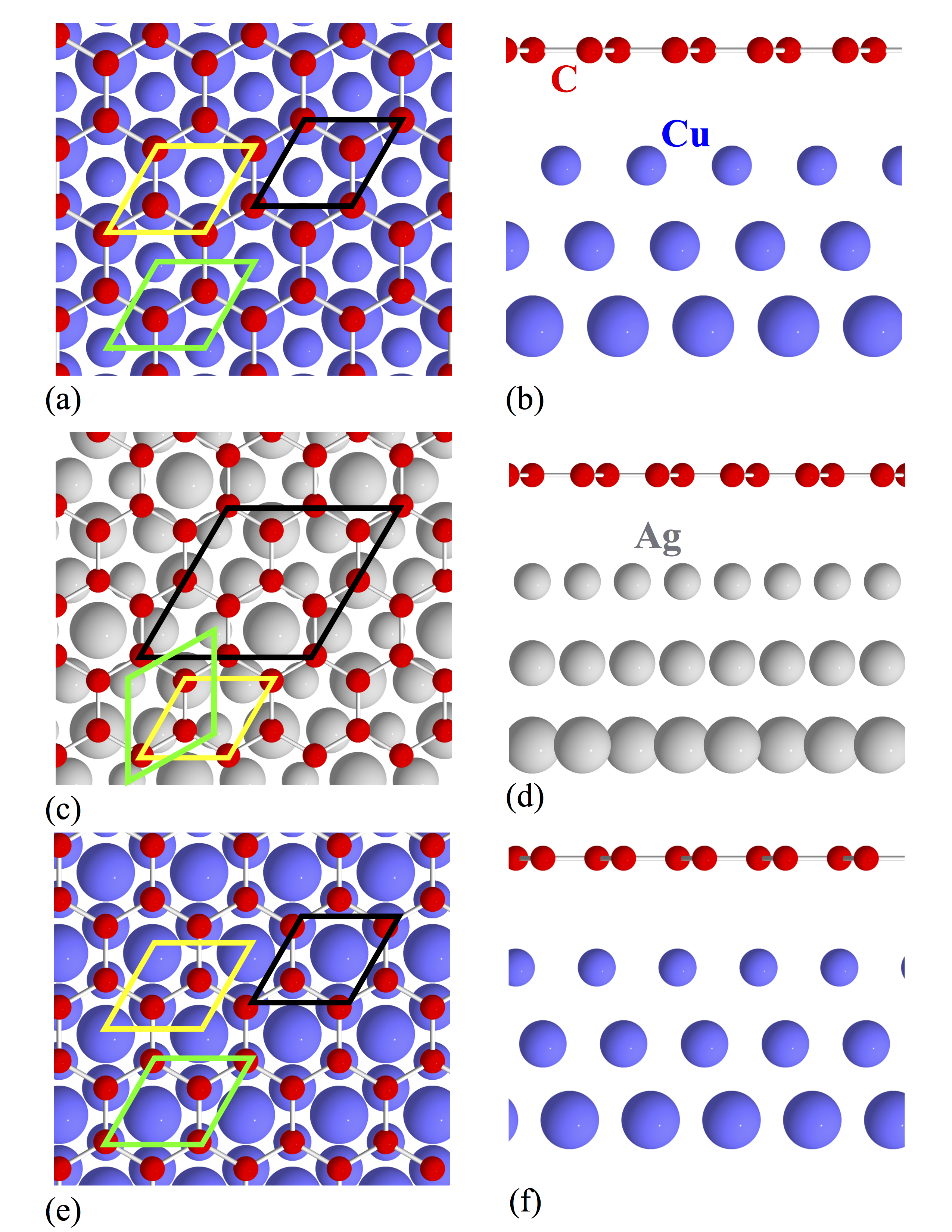}}
  \caption{ \label{fig:graphene_subs} (a) Top view and (b) side view
    of 1:1 match between strained graphene and Cu(111) (or
    Ni(111)). (c) Top view and (d) side view of $2\times2$ graphene
    strained by $<4 \%$ to match the $\sqrt(3)\times\sqrt(3)$
    supercells of Ag(111) (or Pd(111), Pt(111), Au(111) and Al(111))
    surfaces. The primitive cell of graphene is shown with a yellow
    box, that of the substrates with a green box, and lattice-matched
    cells of graphene and substrates with a black box. Another
    configuration for (a) is shown in (e) top view and (f) side view,
    but with a different surface placement of the atoms at the
    interface. The substrate atoms beyond the third layer from the
    interface are hidden for clarity.}
\end{figure}

As an illustrative example, we apply the \textit{MPInterfaces}
framework to determine the lattice-matches of graphene with seven
potential metal substrates. Graphene has a hexagonal lattice with a
lattice parameter of 2.46~\AA. Table~\ref{fig:graphene_subs} lists the
lattice-mismatch between graphene and the (111) surfaces of Pt, Pd,
Cu, Ag, Ni, Au and Al. A 1:1 match between the primitive cells of
graphene and substrate results in lattice-mismatches,
$\epsilon_{1:1}$, exceeding 10\%, except for Cu and
Ni. \textit{MPInterfaces} identifies that a $\sqrt{3}\times\sqrt{3}$
supercell of all the other metal (111) surfaces is matched with the
$2\times2$ surface of graphene within a lattice mismatch,
$\epsilon_{MP} < 4\%$. Figure~\ref{fig:graphene_subs} shows a
schematic of the interfaces obtained for graphene and the seven
substrates and the following code excerpt illustrates how the lattice
matches are obtained in \textit{MPInterfaces}.

\begin{lstlisting}
  # A code excerpt for generation of interfaces 
  # between graphene and substrates 
  # within a give lattice mismatch and maximum
  # area of the interface surface
  # All distances are in Angstroms and lattice 
  # mismatches in percent

  from mpinterfaces.interface import Interface
  from mpinterfaces.transformations import *
  from mpinterfaces.utils import *

  separation = 3 
  nlayers_2d = 2
  nlayers_substrate = 2
  # Lattice matching algorithm parameters
  max_area = 400 
  max_mismatch = 4
  max_angle_diff = 1
  r1r2_tol = 0.01
 
  substrate_bulk = Structure.from_file(
                          'POSCAR_substrate')
  substrate_slab = Interface(substrate_bulk,
                             hkl = [1,1,1],
                             min_thick = 10,
                             min_vac = 25,
                             primitive = False,
                             from_ase = True)
  mat2d_slab = slab_from_file([0,0,1],
                             'POSCAR_2D')
  # get aligned lattices
  substrate_slab_aligned, mat2d_slab_aligned =
         get_aligned_lattices(
               substrate_slab,
               mat2d_slab,
               max_area = max_area,
               max_mismatch = max_mismatch,
               max_angle_diff = max_angle_diff,
               r1r2_tol = r1r2_tol)
  # merge substrate and mat2d in all possible
  # ways
  hetero_interfaces = generate_all_configs(
                        mat2d_slab_aligned,
                        substrate_slab_aligned,
                        nlayers_2d,
                        nlayers_substrate,
                        separation 
\end{lstlisting}  

\subsection{Workflows}
\label{sec:mpint_workflows}

Any computational analysis of a materials system involves an initial
setup of a computational model (generation of the interface models as
described in previous sections), a series of calculations which may
depend on each other and optional automated post processing of the
computed data to extract information and knowledge. This constitutes a
computational workflow.  The \textit{MPInterfaces} package supports the
writing of simple computational workflows where each step is a Python
function that can take as arguments the values returned by the
Python function of the preceding step.

The computational job management and workflow creation extends the
Materials Project packages \textit{Custodian}~\cite{Ong2012b} and
\textit{Fireworks}~\cite{fireworks}. In comparison to the Fireworks
workflows that links together objects of the Fireworks class in a
direct acyclic graph (DAG), our workflows constitute simple linearly
connected jobs described by Python functions.  A major difference
between this workflow and that of Fireworks is that we do not use a
database for the archiving and launching of workflows; this design
aspect of Fireworks enables fetching of the workflows from a remote
database when in need and the subsequent launching of jobs in the
workflow on the cluster. We built on these workflow ideas, and
developed a simple and portable workflow creation and job management
system, that manages computational high throughput discovery projects
run on differing job queue systems like SLURM and PBS.

Our workflows are typically a combination of sequences of structure
input and manipulation, and the calculation of the energy or a
material property of the structure, and post processing of calculated
data to extract required information.  Each workflow is divided into
steps, which can be any part of the aforementioned sequence.  

Our workflow design employs simple Java serializable object notation
(JSON) checkpoint files for keeping track of the jobs and their
completion status on the queue.  The checkpoint files provide a
concise logging of the energies calculated for each calculation in a
single file together with information on input files and parameters
used for each calculation in the workflow.

The checkpoint feature also enables a comprehensive logging of errors
and the handling of specific error measures to ensure succesful
completion of the workflow. This aids in the subsequent analysis of
error sources and directs specific error handling corrections to the
appropriate jobs, where we could not apply generalized auto-correction
schemes. We find the error handling particularly useful in for queue
errors, which can be as high as 10\% on large supercomputing
clusters. We adopted this simplified implementation to avoid the need
for administrative privilege to setup up a dedicated workflow database
on each individual computational resource. Our approach also
simplifies the setting up of the workflow environment on arbitrary
computing resources.

We note and highlight that the job management system caters to the
loads of general investor queues on large supercomputer clusters like
HiPerGator and the XSEDE resources of Stampede which use the PBS and
SLURM job scheduling systems respectively.  In addition, partition
batch jobs management implemented in this high throughput materials
analysis framework facilitates a controlled load on the job scheduler.


To illustrates the features of \textit{MPInterfaces}, we describe in the
following the two computational workflows that create interfaces
structures, perform DFT and MD simulations, respectively, followed by
the subsequent extraction of information.

\subsubsection{VASP workflow example}
\label{sec:mpint_vasp_workflow}

\begin{figure}
\includegraphics[width=6.8cm]{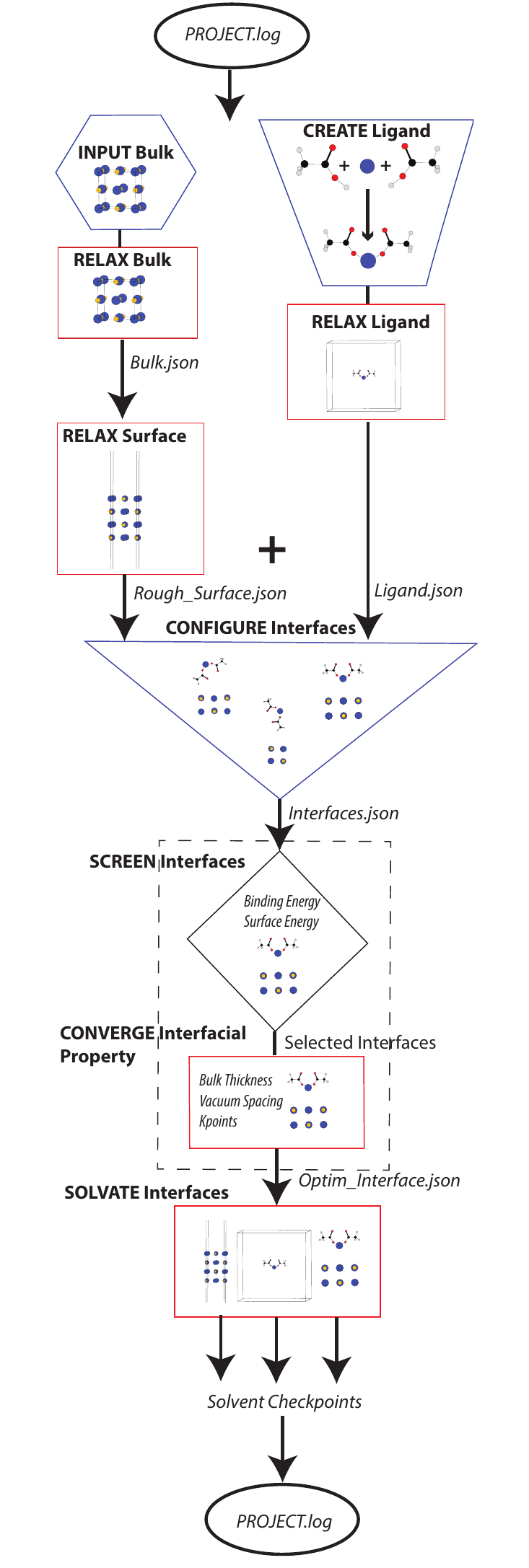}
\caption{ \label{fig:VASP Workflow}
  Workflow to study the adsorption of lead acetate ligands on PbS
  nanocrystal facets using DFT.
  %
    }
\end{figure}

Figure~\ref{fig:VASP Workflow} illustrates an example for a DFT
workflow that employs the VASP code to investigates the ligand
adsorption on nanocrystal facets. The materials' system considered is
the adsorption of lead acetate ligands on PbS facets.

After assembling two acetic acid molecules and a Pb atom to form the
ligand, lead acetate, the first step is to relax the ligand
molecule. The PbS bulk phase is relaxed and slabs are created and
relaxed for the appropriate surface facets. The execution of each step
results in a JSON checkpoint file that is passed on to the subsequent
steps. Combining the information from the ligand and surface slab
relaxation checkpoint files, the next step creates and relaxes ligand
adsorption configurations with different adsorption sites, ligand
orientations, ligand densities {\it etc.} as described in
Section~\ref{sec:mpint_ligand_nano}. The final step extracts the
required interfacial properties such as surface and binding energies
and optionally perform solvation calculations with
VASPsol~\cite{Fishman2013, Mathew2014, Mathew-arxiv2016,
  VASPsol_GitHub} to estimate the effect of solvents or electrolytes
on surface and binding energies.

\subsubsection{LAMMPS workflow example}
\label{sec:mpint_lammps_workflow}

\begin{figure}
\includegraphics[width=7cm]{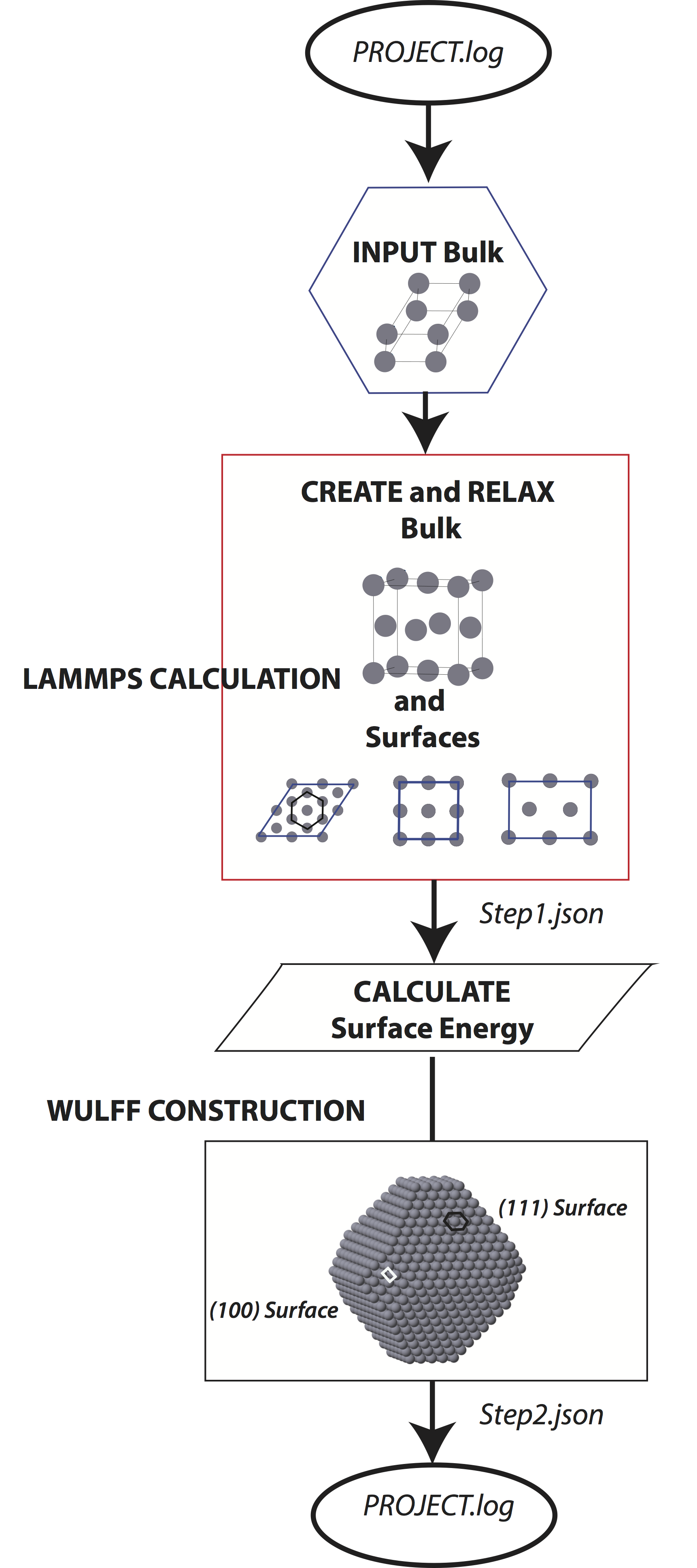}
  \caption{ \label{fig:LAMMPS Workflow}
  Workflow to predict the shape of aluminum nanocrystals using
  empirical energy models with the LAMMPS code.
   }
\end{figure}

\textit{MPInterfaces} also provides an interface to the widely used MD
code, LAMMPS~\cite{Plimpton1995}.  Figure~\ref{fig:LAMMPS Workflow}
illustrate the workflow for the calculation of the surface energies of
Al facets with an empirical energy model implemented into LAMMPS to
construct the shape of Al nanocrystals using the Wulff construction.
The workflow consists of two steps. First, a geometric optimization is
performed for the bulk phase of Al and the constructed low-index Al
facets.  The LAMMPS calculations employ the COMB3
\cite{choudhary2015charge} empirical potential. The second step uses
the energies obtained from the first step to compute the surface
energies and performs a Wulff construction implemented in the
\textit{MPInterfaces} package to create the nanocrystal shape.


\section {Summary}
We have implemented a materials project based open-source Python
package, \textit{MPInterfaces}, that extends the capabilities of
existing high-throughput frameworks such as pymatgen, custodian and
fireworks, towards the study of interfacial systems.  The package is
being continuously developed and is hosted on GitHub at
\url{https://github.com/henniggroup/MPInterfaces}. We demonstrate the
usefulness of the package and its capabilities by various illustrative
examples accompanied by their corresponding code excerpts. Through the
reuse of existing efficient Python tools, with this open-source
undertaking, we intent to provide a collaborative platform that
welcomes any interested user to review and improve the code base and
help push the limits of state of the art computational modeling of
interfaces.

\section{Acknowledgments}
K.\ Mathew and R.\ G.\ Hennig are funded by the National Science
Foundation under the CAREER award No.\ DMR-1056587 and the award
No.\ ACI-1440547, and by the National Institute of Standards and
Technology (NIST) under award 00095176. A.\ Singh is funded by the
Professional Research Experience Postdoctoral Fellowship under award
No.\ 70NANB11H012.  J.\ J.\ Gabriel, F.\ Tavazza and A.\ Davydov are
funded by the Material Genome Initiative funding allocated to NIST.
This research used computational resources provided by the University
of Florida Research Computing (\url{http://researchcomputing.ufl.edu})
and the Texas Advanced Computing Center under Contracts TG-DMR050028N,
TG-DMR140143, and TG-DMR150006.  This work used the Extreme Science
and Engineering Discovery Environment (XSEDE), which is supported by
National Science Foundation grant number ACI-1053575.

\bibliographystyle{elsarticle-num}
\bibliography{References}

\end{document}